\documentclass[12pt, oneside]{article}   	

\usepackage[letterpaper, margin=1in, top=1in, bottom=1in]{geometry}

\usepackage{cite}
\usepackage{nameref}
\usepackage[parfill]{parskip}    		
\usepackage{amssymb}
\usepackage{amsmath}
\usepackage{graphicx}
\usepackage[table,xcdraw]{xcolor}
\usepackage{wrapfig}
\usepackage{float}
\usepackage{caption}
\usepackage{siunitx}
\usepackage{hyperref}
\usepackage{bm}
\hypersetup{
    colorlinks=true,
    linkcolor=blue,
    filecolor=magenta,      
    urlcolor=cyan,
}
\usepackage[T1]{fontenc}
\usepackage{lmodern}
\usepackage[utf8]{inputenc}
\usepackage[]{authblk}

\title{Computing critical exponents in 3D Ising model via pattern recognition/deep learning approach} 

\author[1]{Timothy A. Burt\thanks{taburt@uh.edu}}
\affil[1]{Department of Physics, University of Houston, Houston, TX 77204 USA}

\date{November 4, 2024}

\begin{document}
\eject
\maketitle
\begin{abstract}
In this study, we computed three critical exponents ($\alpha, \beta, \gamma$) for the 3D Ising model with Metropolis Algorithm using Finite-Size Scaling Analysis on six cube length scales (L=20,30,40,60,80,90), and performed a supervised Deep Learning (DL) approach (3D Convolutional Neural Network or CNN) to train a neural network on specific conformations of spin states. We find one can effectively reduce the information in thermodynamic ensemble-averaged quantities vs. reduced temperature t (magnetization per spin $<m>(t)$, specific heat per spin $<c>(t)$, magnetic susceptibility per spin $<\chi>(t)$) to \textit{six} latent classes. We also demonstrate our CNN on a subset of L=20 conformations and achieve a train/test accuracy of 0.92 and 0.6875, respectively. However, more work remains to be done to quantify the feasibility of computing critical exponents from the output class labels (binned $m, c, \chi$) from this approach and interpreting the results from DL models trained on systems in Condensed Matter Physics in general.
\end{abstract}

\eject

\section{Introduction}
In terms of dimensions, the three-dimensional Ising square lattice model is a simple extension of the two-dimensional Ising model, which Onsager solved analytically in 1944. However, to this day, no exact solution exists. Critical exponents are typically found using simulation results. 

This project was primarily motivated by two papers: a computational physics simulation study and a pattern recognition study using DL on a 2D Ising model. 

Sonsin \textit{et al.} \cite{sonsinComputationalAnalysis3D2015} simulated the 3D Ising system using a Monte Carlo method/Metropolis algorithm and found that for three system sizes of L=150,200,250, the Metropolis approach (local spin update) can reach equilibrium in a feasible amount of time (10000 Monte Carlo Sweeps (MCS)) and found the critical temperature $T_c$ to be $\approx 4.512$ $k_B T/J$ from Binder's Cumulant $U_4=1-\frac{<m^4>}{3<m^2>^2}$. Their approach contrasts the standard global spin update methods typically used for large system sizes of L (e.g., Wolff, Swendsen-Wang algorithms), which avoid the critical slowing down near $T_C$.

Holzbeck \textit{et al.} \cite{holzbeckPatternRecognition2DIsing2019} investigated the phase transition of the 2D Ising model using a DL approach to recognize patterns in spin conformations from equilibrated Monte Carlo Metropolis algorithm simulations. They found the trained DL model could distinguish between conformations with an inner region coupling constant $J_{inner}=2$ from conformations with an outer region coupling constant $J_{outer}=1$.

The two separate but primary goals for this project were to:
\begin{enumerate}
\item Compute three critical exponents for the 3D Ising model using finite-size scaling analysis (FSSA) on simulation results, replicating the model and parameters in \cite{sonsinComputationalAnalysis3D2015}.
\item Design, implement, and evaluate a supervised Deep Learning (DL) approach to recognize specific realizations of spin states, each categorized by a unique pattern or montage, extending the work from \cite{holzbeckPatternRecognition2DIsing2019}.
\end{enumerate}

This paper was initially written as a final project for the PHYS 7350 Advanced Computational Physics class at the University of Houston (UH) in Spring 2020.

\section{Methods}

For this study, we design a method for determining critical exponents from specific state realizations labeled with a class corresponding to three phases of matter using a supervised DL approach. We outline each step in our scheme below:

\begin{enumerate}
\item Run 3D Ising model (ferromagnetic square lattice PBC, Metropolis algorithm) simulations
 \begin{itemize}
 \item Input
  \begin{itemize}
 \item Parameters: T=[3.5,5.5], ($k_B$ \& J=1), $\Delta T=0.05$, Total Monte Carlo Sweeps (MCS): 10000, L=20,30,40,60,80,90, \textbf{H}=0, 2 independent runs (different seed, both all spins up initially) at each L,T value \item Random number seeds used: [2,9]\footnote{Mersenne Twister algorithm, mt19937 (C++11)}
 \item The Hamiltonian used is shown in Equation \ref{eq:model_H}
 \end{itemize}
 \item Output
 \begin{itemize}
 \item Trajectories: MC sweep (MCS), E(t), M(t) (and running averages of each)
 \item Conformations: output every 25 MCS (only use $MCS\geq6500$ frames, based on magnetization equilibration / correlation time $<\tau_m>+<\tau_m>_{\sigma}$
 \end{itemize}
 \item Number of conformations saved: 282\footnote{(41 T's * 6 L's * 3500 equilibrated realizations * 2 independent runs (different seed, both all spins up initially)) / 25 (stride of realization/trajectory output)=282 conforms} conforms available at each L,T value
\item Train/test ratio: 70/30
\item Using that ratio gives 19,740 realizations for training and 8,460 for testing, randomly drawn from 6 predefined latent classes (over L,T) evenly (to prevent overfitting/bias)
 \end{itemize}
 \item Calculate correlation (or equilibration times) averaged over both runs for all L,T
\item Calculate thermodynamic averaged c(T), m(T), $\chi$(T) values
 \begin{itemize}
\item Bin each of these plots into six latent classes by manually identifying three ranges for each of the three plots (only 6 out of 27 possible classes had non-zero bins)
 \end{itemize}
 \item Run Finite-Size Scaling Analysis (FSSA) to compute critical exponents c,m,$\chi$ on two sets of L ranges, L=20,40,80 \& L=30,60,90. This step develops intuition about simulation accuracy by comparing these values to the literature.
\item Preprocess data and prepare it for DL algorithm training
 \begin{itemize}
\item Loading conformation sparse .txt files and compressing/cutting unused frames
\item Train/test split over six classes (evenly)
 \end{itemize}
\item 3D CNN DL training/testing\footnote{Implemented in Python and TensorFlow v1}
 \begin{itemize}
\item Input: Spin conformation only
\item Output: Latent class (6 categories)
 \end{itemize}
\item Run FSSA \textbf{again} to compute critical exponents c,m,$\chi$ on train/test sets (mixed L's) independently. The final accuracy of the DL algorithm is based on the percent difference between the two sets.
\end{enumerate}

\begin{equation}
\begin{split}
H=-J\sum_{i,j,k=1}^{N}( S_{i-1,j}S_{i,j} + S_{i,j}S_{i+1,j} + S_{i,j-1}S_{i,j} + S_{i,j}S_{i,j+1} + S_{i-1,k}S_{i,k} + S_{i,k}S_{i+1,k} \\
+ S_{i,k-1}S_{i,k} + S_{i,k}S_{i,k+1} + S_{j-1,k}S_{j,k} + S_{j,k}S_{j+1,k} + S_{j,k}S_{j,k-1} + S_{j,k}S_{j,k+1})-\bm{H}\sum_{i}^{N}\bm{s_i}
\end{split}
\label{eq:model_H}
\end{equation}

\section{Results/Discussion}

\subsection{Simulation and finite-size scaling analysis}

\subsubsection{Equilibration times}

\begin{figure}[H]
\centering
\captionsetup{justification=centering}
\includegraphics[width=\linewidth]{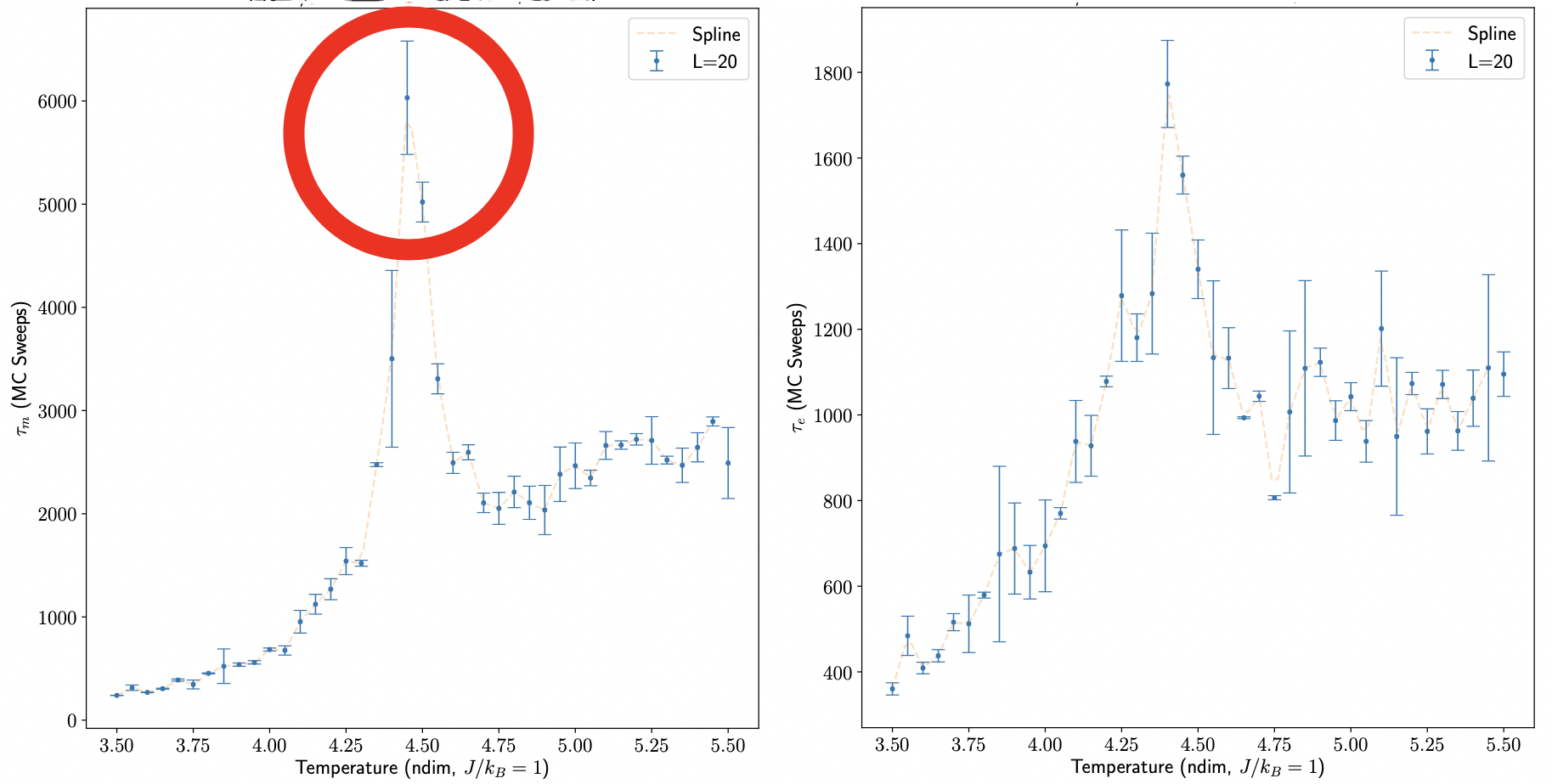}
\caption{Equilibration times for L=20 subset. Magnetization per spin $<\tau_m>$ left, energy per spin $<\tau_e>$ right. Two independent runs were concatenated together to give the ensemble mean/std. err.}
\label{fig:equil_times}
\end{figure}

Figure \ref{fig:equil_times} shows magnetization per spin $<\tau_m>$ and energy per spin $<\tau_e>$ equilibration times vs. temperature T for L=20. Since L=20 near $T_C$ had the largest overall L values, only conformations with $MCS\geq6500$ were used as training/testing data (came from $<\tau_m>+\sigma <\tau_m>_{\mu}$). 

Note that throughout the paper, the ensemble average ($<\cdot>$) is defined over the two independent simulation runs.

\subsubsection{\texorpdfstring{$<c>(T), <m>(T), <\chi>(T)$}{<c>(T), <m>(T), χ(T)} results}

\begin{figure}[H]
\centering
\captionsetup{justification=centering}
\includegraphics[width=\linewidth]{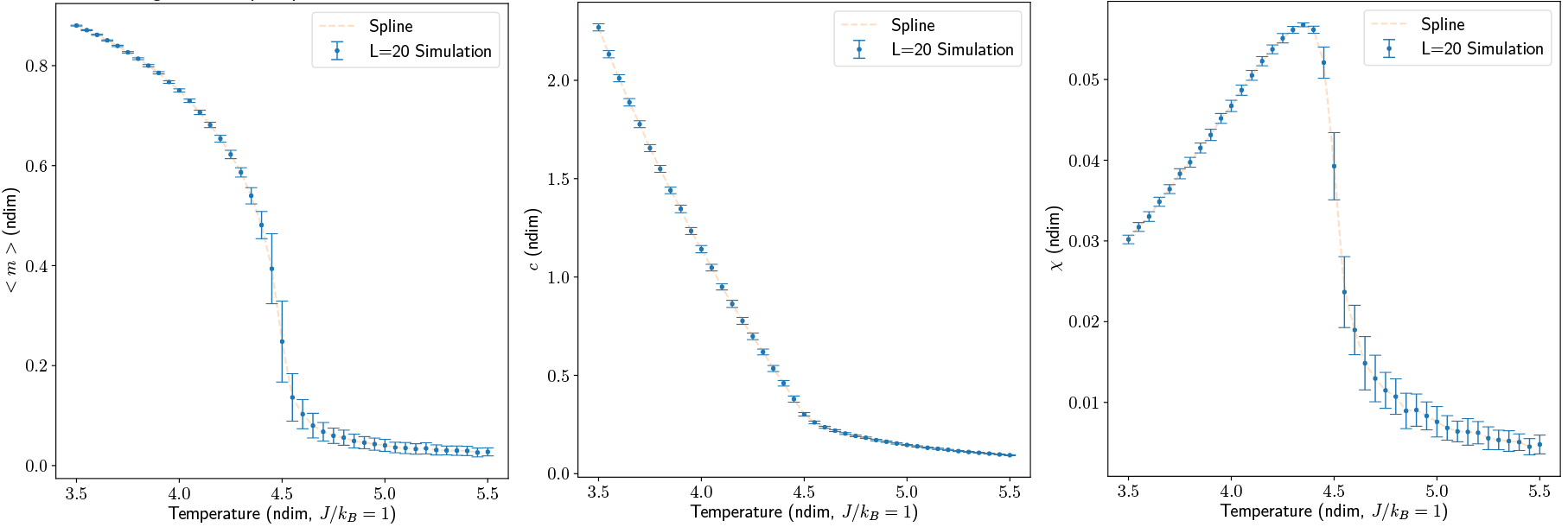}
\caption{From left to right -- Ensemble-averaged $<m>(T), <c>(T), <\chi>(T)$ for L=20. Error bars are $\pm\sigma_{\mu}$. A bootstrapping technique was used to generate fluctuation statistics $(c, \chi)$.}
\label{fig:mcchi_vs_T}
\end{figure}

Figure \ref{fig:mcchi_vs_T} shows ensemble averaged $<c>(T), <m>(T), <\chi>(T)$ values for L=20 (other L results similar). The statistics are sufficiently sampled to estimate the corresponding critical exponents $\alpha, \beta, \gamma$.

We see some key differences between the 2D and 3D systems. First, the magnetic susceptibility and specific heat per spin are much more complex to sample well in 3D due to their system size-dependent fluctuations (using the Metropolis algorithm). Second, strong finite-size effects, especially for m near $T_C$, will cause significant errors in the small system sizes and range used.

\subsubsection{Critical exponents \texorpdfstring{$\alpha, \beta, \gamma$}{α, β, γ} results from FSSA}

\begin{figure}[H]
\centering
% \captionsetup{justification=centering}
\captionsetup{justification=raggedright,singlelinecheck=false,format=hang}
\includegraphics[width=\linewidth]{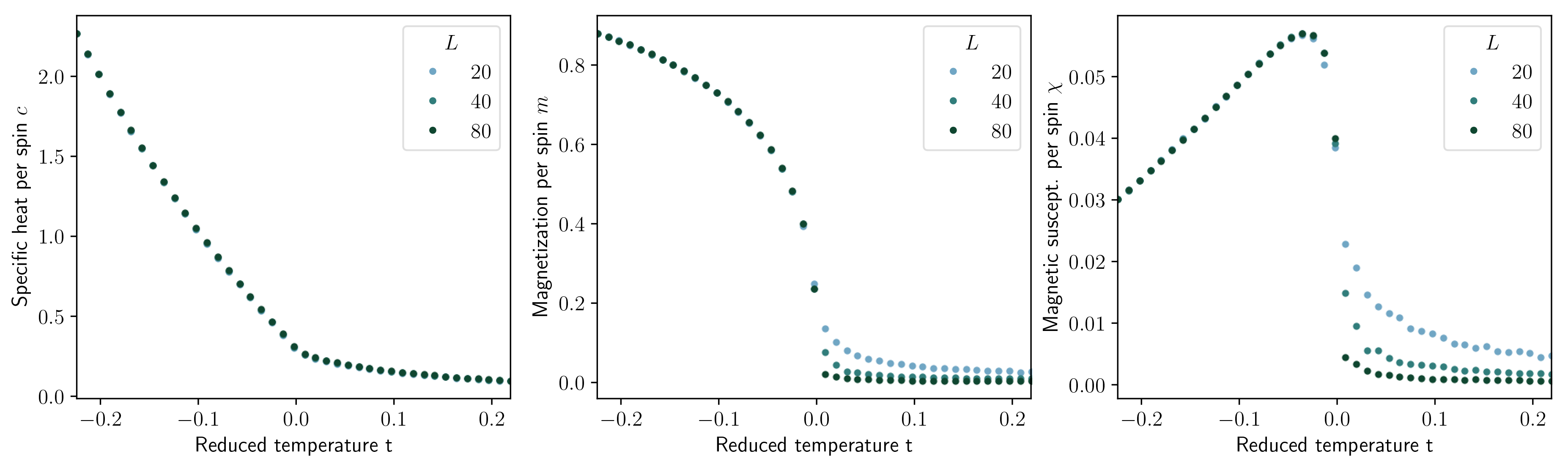}
\includegraphics[width=\linewidth]{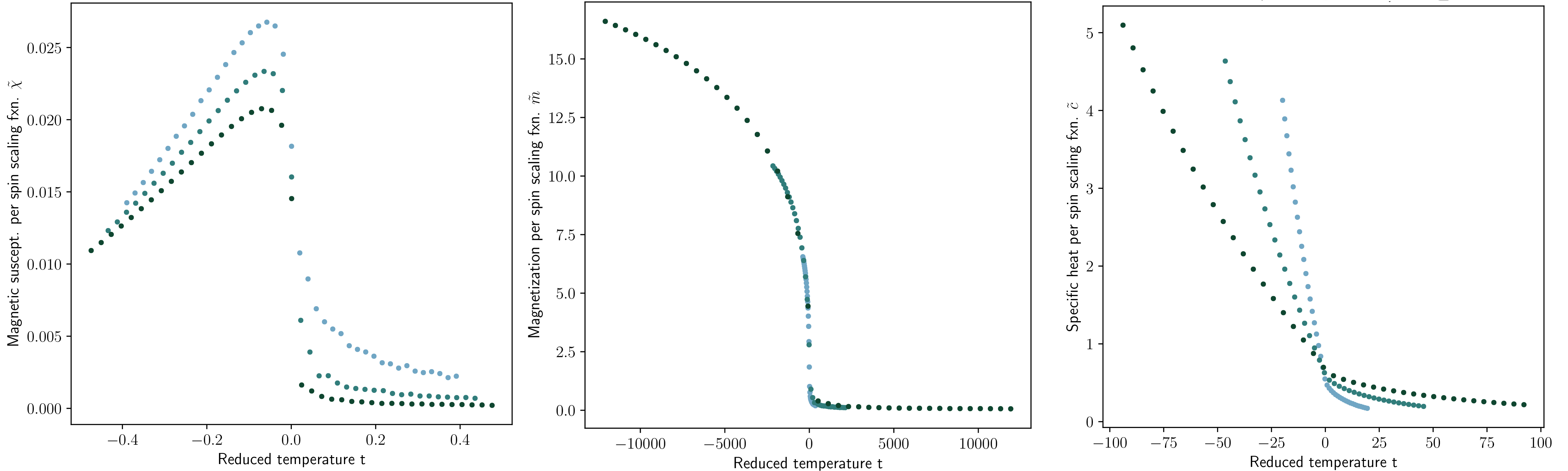}
\caption{Top left to right -- Uncollapsed results for $c, m, \chi$ vs. reduced temperature $t\equiv \frac{T-T_C}{T_C}$ on L=20,40,80 set (L=30,60,90 not shown).\\\vspace{1mm}
Bottom left to right -- Corresponding data collapsed scaling functions for $\tilde{c}, \tilde{m}, \tilde{\chi}$ after FSSA was performed, using the determined critical exponents.}
\label{fig:fss_results}
\end{figure}

Figure \ref{fig:fss_results} shows the uncollapsed and the corresponding data collapse of the scaling functions for $c, m, \chi$ vs. reduced temperature on the L values L=20,40,80 (L=30,60,90 similar, not shown). The Python package \textit{pyfssa} \cite{sorgePyfssa0762015} was used to perform automated parameter tuning to find the best fits for the critical exponents $\alpha, \beta$. Let us introduce the scaling function conventions used (from \cite{newmanAnalysingMonteCarlo1999}).

\begin{equation}
c=L^{\alpha/\nu}\tilde{c}(L^{1/\nu}t)
\end{equation}

\begin{equation}
m=L^{-\beta/\nu}\tilde{m}(L^{1/\nu}t)
\end{equation}

\begin{equation}
\chi=L^{\gamma/\nu}\tilde{\chi}(L^{1/\nu}t)
\end{equation}

Since the exponents are related and only two are independent, we can reduce these fit values in the following way:
\begin{enumerate}
    \item Obtain $\alpha, \delta \alpha, \nu, \delta \nu$ directly from \textit{pyfssa} on c (using a logarithmic power law $\ln(L)$ near $T_C$).
    \item Obtain $\beta, \delta \beta, \nu, \delta \nu$ directly from \textit{pyfssa} on m.
    \item $\gamma$ was determined from the $\alpha, \beta$ fit values using the scaling constraint $\alpha + 2\beta + \gamma = 2$ \cite{rushbrookeThermodynamicsCriticalRegion1963}, propagating the errors. $\chi$ was collapsed using a logarithmic power law $\ln(L)$ near $T_C$.
\end{enumerate}

The rationale for fitting c by substituting $\ln(L)$ into the scaling function comes from \cite{kaupuzsEnergyFluctuationsSingularity2004}, which stated the singularity of specific heat near criticality for the 3D Ising system scales like a logarithmic-law from MC simulation results, not a power-law (this differs from the mean-field theory prediction).

Table \ref{tab:fss_results} shows the determined exponents with error, along with a comparison to literature values. For $\alpha, \beta$, errors come from the \textit{pyfssa} output. For $\gamma$, the error is determined by propagating the errors in the equation $\alpha + 2\beta + \gamma = 2$, using $\delta \alpha,\delta \beta$ from \textit{pyfssa}. The results seem close to the values of the literature despite the challenges described by the small system sizes and ranges used. Notice how c vs. reduced temperature t hardly scales over this region. 

However, after a closer analysis, it is clear that the actual errors on the critical exponents from data collapse are highly underestimated due to systematic errors propagated within the pipeline. 

First, the fact that errors are greater for the larger system size set with all other parameters held constant is a clear sign the \textit{pyfssa} determined errors on $\alpha, \beta$ are shifted by a constant bias, introduced by setting the initial guesses for the critical reduced temperature, critical exponent and the corresponding correlation length critical exponents $\rho_0, \zeta_0, \nu_0$, to the accepted values. The results are highly dependent on the initial guess provided. 

These errors could be accounted for by first doing manual scaling and data collapse on the three tuning parameters, giving the critical exponent an upper and lower bound. Next, we would run the \textit{autoscale} function in the package, using the manually determined three parameters as our initial guesses, along with errors for each data point shown at the top of Figure \ref{fig:fss_results}. The error values then output from the function would still account only for random errors. 

To incorporate systematic errors, which are the error bounds found during manual scaling, we can use error propagation (assuming upper/lower errors are uniform) to get the total error. The random errors could be weighted by the quality of data collapse S, defined as the reduced $\chi^2$ statistic, which is output from the package (but was not used here). The systematic errors would have a weight of 1 in the error propagation.

Another paper utilizing a similar technique to approximate the critical exponents for the 3D Ising system provided results of $\beta=0.18\pm0.02$, $\gamma=1.0\pm0.1$, $\alpha=0.10\pm0.02$ \cite{delgenioAnomalousOrderingInhomogeneously2010}.

\begin{table}[H]
\centering
\captionsetup{justification=centering}
\resizebox{\textwidth}{!}{%
\begin{tabular}{|l|l|l|
>{\columncolor[HTML]{FFFFFF}}l |l|}
\hline
\cellcolor[HTML]{000000} & L=20,40,80 results                             & L=30,60,90 results                             & Accepted results \cite{UniversalityClass2024}        & Uncertainty range ($1\sigma$ confidence) (random errors)                             \\ \hline
$\alpha$                 & $0.134 \pm 0.078$                              & $0.084 \pm 0.075$                              & $0.11008 \pm 0.00001$   & 21.7-31.0 \%                                    \\ \hline
$\beta$                  & $0.269 \pm 0.047$                              & $0.248 \pm 0.032$                              & $0.326419 \pm 0.000003$ & 17.6-24.0 \%                                    \\ \hline
$\gamma$                 & $1.327\pm 0.123$                               & $1.420 \pm 0.098$                              & $1.237075 \pm 0.000010$ & 7.27-14.8 \%                                    \\ \hline
$T_C$ (ndim)             & \cellcolor[HTML]{B9B9B9}\textless{}0.1\% error & \cellcolor[HTML]{B9B9B9}\textless{}0.1\% error & 4.511523                & \cellcolor[HTML]{000000}{\color[HTML]{000000} } \\ \hline
\end{tabular}%
}
\caption{Measured critical exponents $\alpha, \beta, \gamma$ from simulation for sets L=20,40,80 \& L=30,60,90 compared with accepted values and an uncertainty range for the random errors (determined using percent error formula on each exponent), along with the accepted value for $T_C$.}
\label{tab:fss_results}
\end{table}

\subsection{Using Deep Learning on spin states to predict different phases of matter}

Since a standard CNN was implemented in this paper, it was intended for classification tasks (i.e., supervised learning). Classification ML models only output latent class labels with no uncertainty, while regression ML models can output continuous values (i.e., unsupervised learning). This difference is due to a typical CNN algorithm's standard dropout layer after the fully connected layers. Therefore, we must assign categories to ranges of the continuous quantities we want to predict.

To devise a methodology for choosing enough bins to give accurate distributions of $m, c, \chi$ while ensuring they are physical and well-defined is a trial and error process. We could also consider this as \textit{course-graining} the physics.

\subsubsection{Class label definitions and interpretation}

\begin{figure}[H]
\centering
% \captionsetup{justification=centering}
\captionsetup{justification=raggedright,singlelinecheck=false,format=hang}
\includegraphics[width=\linewidth]{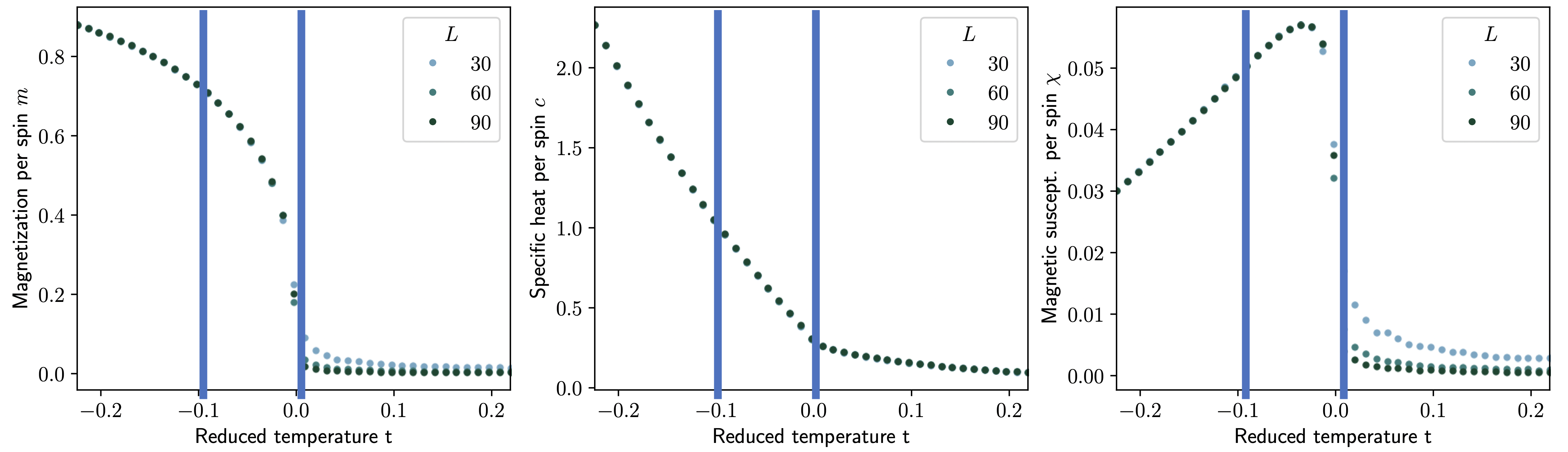}
\includegraphics[width=\linewidth]{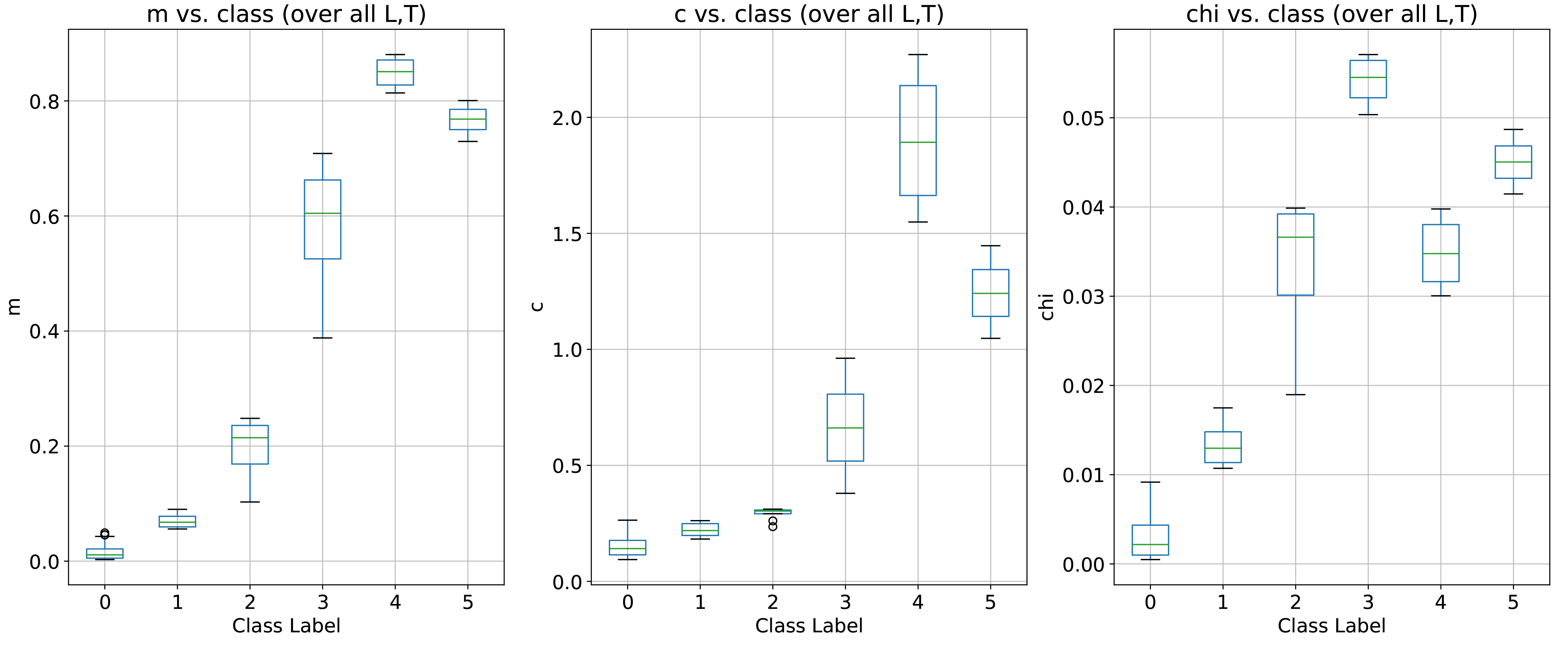}
\caption{Top -- Combined results for $m, c, \chi$ vs. t for L=30,60,90.\\\vspace{1mm}
Bottom -- Box \& whisker plot depicting each quantity split into each of the six classes and its resulting statistics.}
\label{fig:class_bins}
\end{figure}

Let us define three constant ranges (bin edges) on the reduced temperature t. Since there are three quantities we aim to predict ($m, c, \chi$), there are three ranges for each of the three quantities (Figure \ref{fig:class_bins}).

\begin{itemize}
\item $t<-0.1$: Ferromagnetic region
\item $-0.1 \leq t < 0.01$: Transition region
\item $t \geq 0.01$: Paramagnetic region
\end{itemize}

Combining all of the possibilities results in 27 possible latent classes. The Python package \textit{pandas} was used to analyze \& bin the data. After performing standard techniques from statistics, we find there are only six non-empty categories (i.e., for each L,T value, only 6 combinations of the 27 possible bins exist). With the amount of data available, we should be able to adequately train our DL model to detect which class a given conformation belongs to using six classes.

Table \ref{tab:class_dl_info} interprets the six defined class labels. By interpreting the box \& whisker plot in Figure \ref{fig:class_bins}, one can see how each class maps to a region of the phase diagram.

\begin{table}[H]
\centering
\captionsetup{justification=centering}
\resizebox{\textwidth}{!}{%
\begin{tabular}{|c|c|c|c|c|c|c|c|}
\hline
Class Label    & Class Description              & m bin                    & c bin                    & $\chi$ bin               & Train Count & Test Count & Total Count \\ \hline
0              & Far $T_C$ paramagnetic phase   & (-0.001, 0.1{]}          & (-0.001, 0.37{]}         & (-0.001, 0.01{]}         & 9           & 5          & 14          \\ \hline
1              & Near $T_C$ paramagnetic phase  & (0.1, 0.72{]}            & (0.37, 1.04{]}           & (0.04, 0.0568{]}         & 5           & 3          & 8           \\ \hline
2              & $T>T_C$ transition phase       & (0.72, 0.88{]}           & (1.04, 2.269{]}          & (0.01, 0.04{]}           & 4           & 3          & 7           \\ \hline
3              & $T<T_C$ transition phase       & (0.72, 0.88{]}           & (1.04, 2.269{]}          & (0.04, 0.0568{]}         & 3           & 2          & 5           \\ \hline
4              & Near $T_C$ ferromagnetic phase & (-0.001, 0.1{]}          & (-0.001, 0.37{]}         & (0.01, 0.04{]}           & 2           & 2          & 4           \\ \hline
5              & Far $T_C$ ferromagnetic phase  & (0.1, 0.72{]}            & (-0.001, 0.37{]}         & (0.01, 0.04{]}           & 2           & 1          & 3           \\ \hline
Miniset Totals & \cellcolor[HTML]{000000}       & \cellcolor[HTML]{000000} & \cellcolor[HTML]{000000} & \cellcolor[HTML]{000000} & 25          & 16         & 41          \\ \hline
\end{tabular}%
}
\caption{Physical interpretation of class labels in terms of 3 spin phases and 2 regions within each phase, along with the bin ranges for $m, c, \chi$. Also given is the number of each class conformations used in the L=20 miniset, which defines the train/test batch size. The total count is 41 (the number of T values available per L).}
\label{tab:class_dl_info}
\end{table}

\subsubsection{DL implementation and results}

Due to time constraints, we trained/tested on a small subset of L=20 conformations (defined as the \textbf{miniset}). For a detailed description of the DL model, please see \nameref{sec:appendix}.

The input size to the CNN is 256x256x40 (width,height,depth), respectively. This input size can handle any L size from our simulations, which can be rotated to fit inside those dimensions. Extra space not filled with a conformation, e.g., L=20, is padded with zeros, left unscaled, and the conformation repositioned if needed to fit inside.

To reduce the chance of overfitting the model, we split each of the six classes into their own train/test set using \textit{pandas DataFrame.shuffle} method. Given the conformation data, the RandomState seed passed was 6; one can use this same value to reproduce our exact train/test sets. 

We also shuffled each batch of training conformations randomly each time so that the neural network (NN) saw the different classes as out of order. This choice increases the robustness of the trained model for distinguishing between different conformation classes.

The model is trained on conformations for 20 epochs (20 passes over 5 batches of 25 realizations, with accuracy computed every epoch (after 125 realizations, the size of the L=20 miniset). The training batch size is 125, which defines one epoch (when the neural network weights are updated with probability 0.5). 

At this point, the accuracy of the final training is computed. After training, the test accuracy was measured on a separate miniset of 16 realizations that had not been used during training. A batch's train/test accuracy is defined as \#correct/batch size (sometimes called the test sensitivity).

\begin{figure}[H]
\centering
% \captionsetup{justification=centering}
\captionsetup{justification=raggedright,singlelinecheck=false,format=hang}
\includegraphics[width=1.0\linewidth]{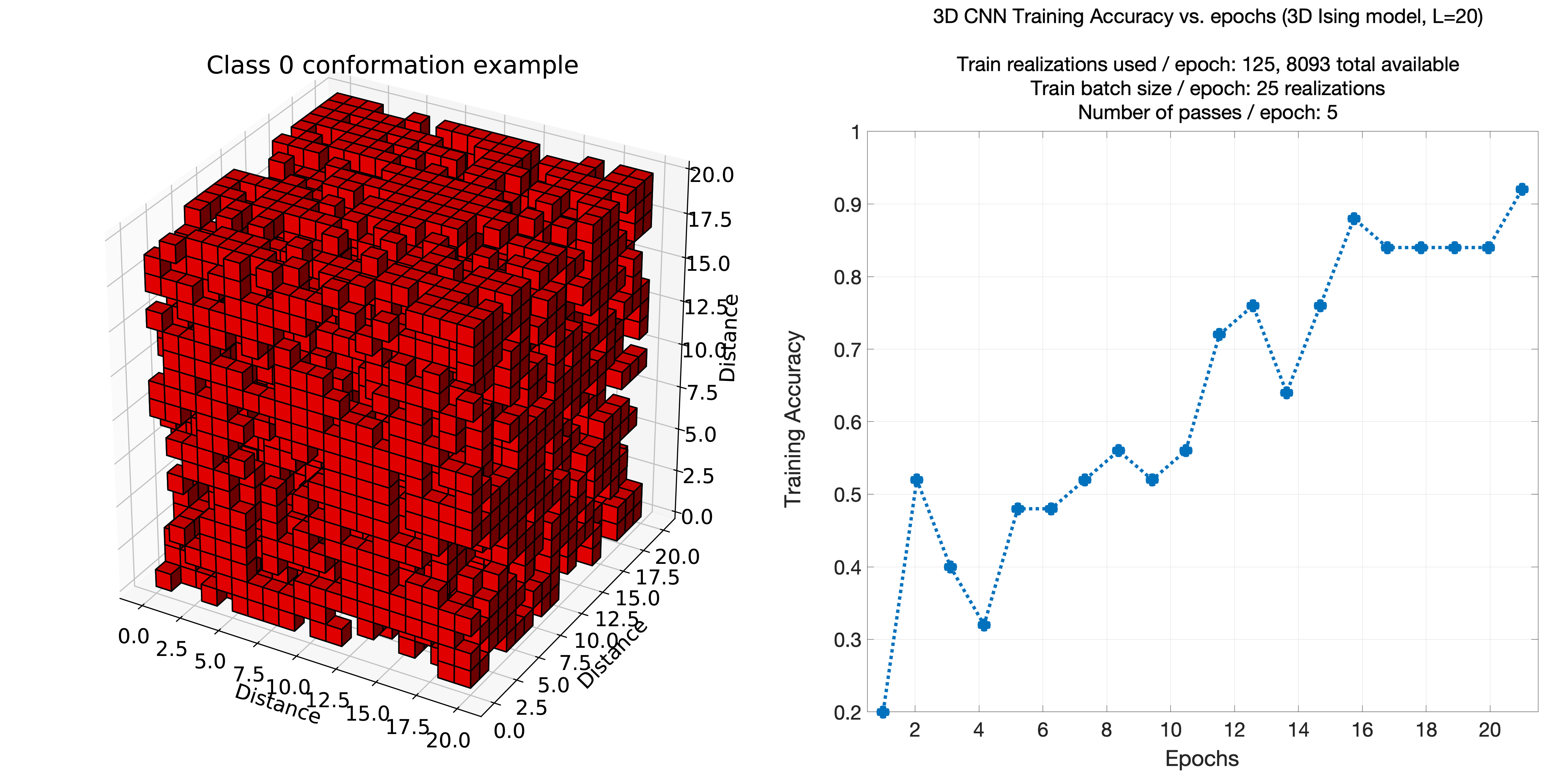}
\caption{Left -- Class 0 (Far $T_C$ paramagnetic phase) L=20 conformation example, before padding. Red voxels represent spin up (1), empty voxels represent spin down (-1).\\\vspace{1mm}
Right -- Training accuracy vs. epochs on the L=20 miniset.}
\label{fig:cf_train_acc}
\end{figure}

We were able to successfully train our model using this subset of L=20 conformations ($125/8093 \approx 1.5$\% of total available (Table \ref{tab:dl_train_test_acc})). Figure \ref{fig:cf_train_acc} shows a plot of training accuracy vs. epochs. 

Unfortunately, we did not split each batch evenly among the classes due to time constraints, which likely caused the ferromagnetic state classes (4,5) to be undersampled and more challenging to detect. This decision explains the lower testing accuracy than the final training accuracy.

\begin{table}[H]
\centering
\captionsetup{justification=centering}
\resizebox{0.25\textwidth}{!}{%
\begin{tabular}{|c|c|}
\hline
Train acc. & Test acc. \\ \hline
0.92                    & 0.6875        \\ \hline
\end{tabular}%
}
\caption{Final training accuracy (left) and test accuracy (right) of the DL model on the L=20 miniset. Training time for the miniset, over 20 epochs, was 5.76 hours on a 2019 iMac running in serial.}
\label{tab:dl_train_test_acc}
\end{table}

\begin{figure}[H]
\centering
\captionsetup{justification=centering}
\includegraphics[width=1.0\linewidth]{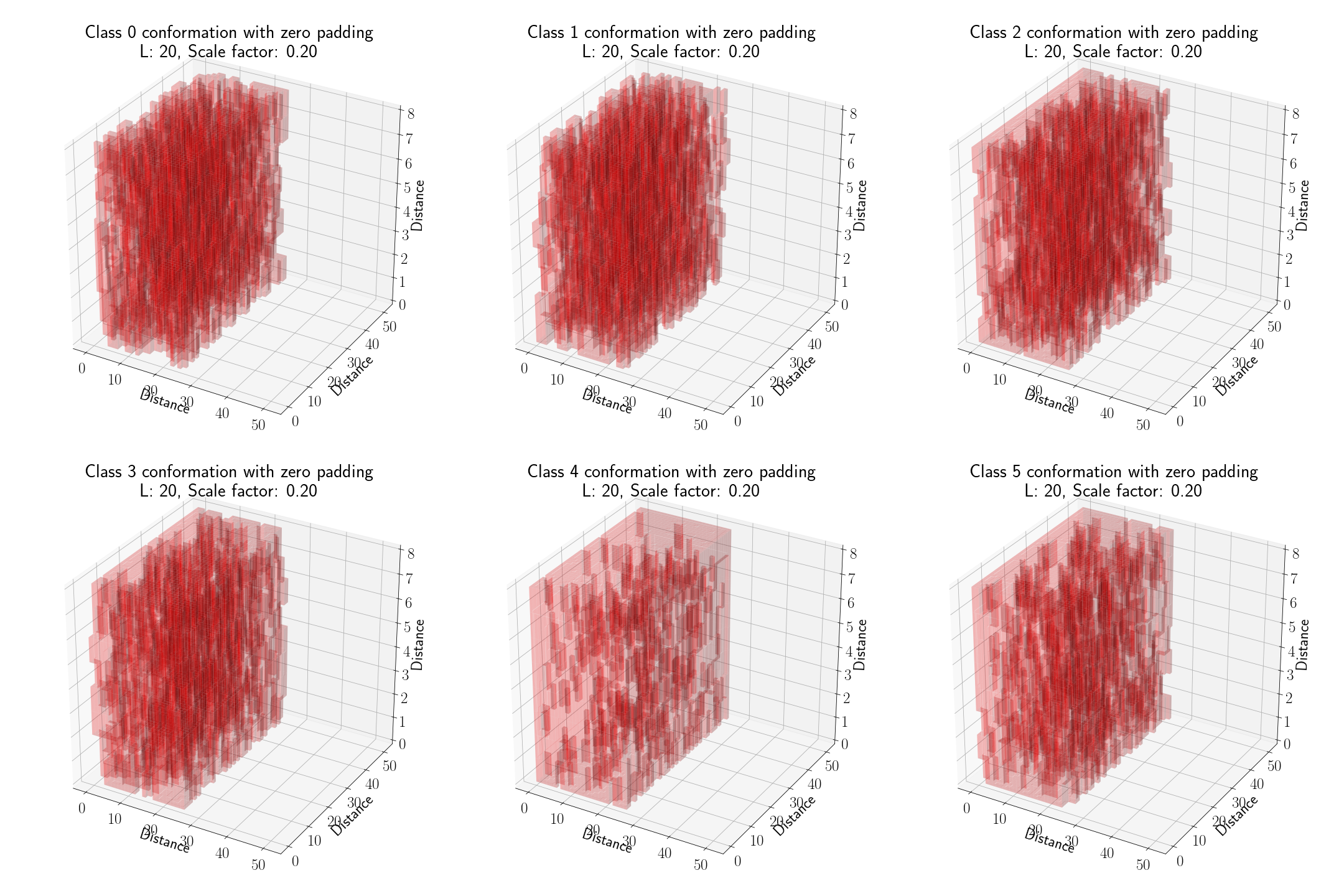}
\caption{Visualizations of all six class conformations, after zero padding to fit the L=20 cube into the 256x256x40 input required for the CNN. Note that the plots are uniformly scaled to 20\% of the actual size for computational efficiency. Padding is added in the depth channel (size 40) along the vertical plane in these diagrams.}
\label{fig:padded_cfs}
\end{figure}

A visualization of all six class example states, as seen by the CNN, with zero padding, is given in Figure \ref{fig:padded_cfs}.

\section{Conclusions}

We have partially demonstrated a method based on intuition for training a 3D CNN to distinguish between 6 latent classes of spin realizations on a miniset of L=20 spin states. We also computed three critical exponents $\alpha, \beta, \gamma$ for the 3D Ising model from Monte Carlo Metropolis simulations using FSSA on two sets of L values separately and provide a pipeline for measuring and determining the accuracy of critical exponents output by the trained DL model.

We computed the error between our measured critical exponents and literature and found systematic biases to be why our reported errors were too small. We then give a procedure that should correctly account for both random and systematic errors. Two ways to improve our critical exponent values would be to use larger system sizes, spanning a range of 1-2 orders of magnitude (10-100x), along with checking the literature for empirical formulas for corrections to $m, c, \chi$ for small L values to better collapse the data.

While the test accuracy was lower than expected, we identified not using a batch size with equally distributed class conformations as the likely culprit. Since both train and test accuracy are higher than chance by a fair amount, we can conclude that the model found features that characterize most of the spin conformations.

While we still need to implement the method entirely for our system, we show that the DL model can learn the six classes for one L value with high accuracy. Therefore, this can be extended to all six L values by using more computational time and resources to train the DL model.

Implementing our method on the 3D Ising model is a significant amount of work, and some might question why one should bother with a data-driven approach instead of the traditional finite-size scaling ansatz to measure critical exponents in condensed matter systems. Our method could be used to classify states of significantly more complicated systems (e.g., systems utilizing Landau-Ginzberg Free Energy Functionals) where identifying new, rarely sampled states needs to be identified in real-time or in systems where the finite-size scaling ansatz is not known beforehand.

\section*{Acknowledgements}
I want to thank Prof. Kevin E. Bassler for mentoring the class project and instructing the PHYS 7350 course at UH. I also thank Prof. Ioannis A. Kakadiaris for providing feedback on the paper for a presentation at TSAPS 2020.

\bibliography{manuscript}

\newpage

\section*{Appendix}
\label{sec:appendix}

\subsection*{CNN architecture}
\label{sec:cnn_arch}

Outlined here are the exact layers in the implemented 3D DL architecture. It is a \textit{vanilla} 3D CNN implemented serially using TensorFlow v1 in Python. The training step uses the Adam Optimizer to minimize the cross entropy with a constant learning rate of $\num{1e-3}$. Specific hyperparameters for the L=20 miniset used are given in Figure \ref{fig:cf_train_acc}. The layers are described below.

\begin{itemize}
 \item First Convolutional Layer: 32 features for each 5x5 patch (1 feature - 32 features)
 \item First Max Pool Layer: max pooling over 2x2 blocks (output image: 14x14x32)
 \item Second Convolutional Layer: 64 features for each 5x5 patch (32 features - 64 features)
 \item Second Max Pool Layer: max pooling over 2x2 blocks (output image: 7x7x64)
 \item First Fully Connected (Densely Connected) Layer: 1024 neurons to process the entire image (output: 1024)
 \item Second Fully Connected (Readout) Layer: output size 6 (one per class), e.g., [1 0 0 0 0 0] would be class 0 output
\end{itemize}
 
\begin{figure}[H]
\centering
\captionsetup{justification=centering}
\includegraphics[width=\linewidth]{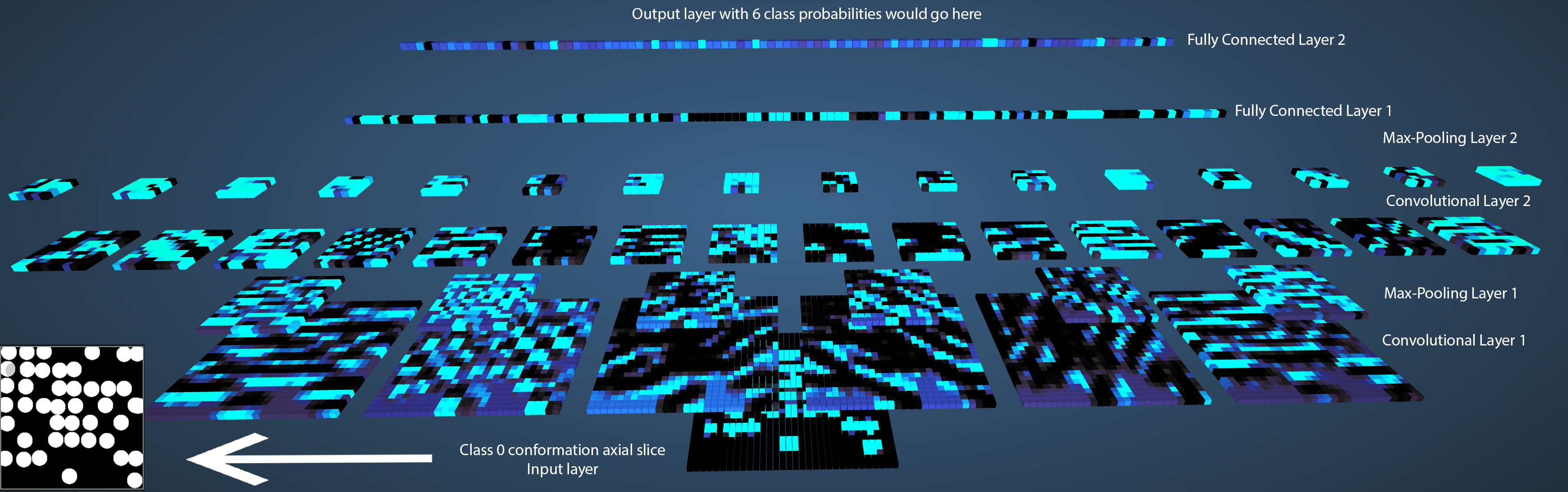}
\caption{Example depiction of a 2D CNN architecture that has the same structure as the one used in this paper (but with one less dimension), with an input layer of a Class 0 (Far $T_C$ paramagnetic phase) axial slice. Generated using \cite{harleyInteractiveNodeLinkVisualization2015}.
}
\label{fig:cnn_layers_ex}
\end{figure}

A depiction of the features each layer of a 2D implementation of our CNN finds on a Class 0 (Far $T_C$ paramagnetic phase) spin state axial slice is shown in Figure \ref{fig:cnn_layers_ex}.

\end{document}